\documentclass[twocolumn,amssymb,amsfonts,floatfix]{revtex4}
\usepackage{bm}
\usepackage{dcolumn}
\usepackage{bm}
\usepackage{mathbbol}
\usepackage{graphicx}
\usepackage{epstopdf} 
\usepackage{amsmath,amsthm}
\usepackage{times}
\usepackage{color}
\usepackage{lipsum}

\begin{document}

\title{Inevitable power-law behavior of isolated many-body quantum systems and how it anticipates thermalization}

\author{Marco T\'avora$^1$, E. J. Torres-Herrera$^2$, and Lea F. Santos$^1$}
\affiliation{$^1$Department of Physics, Yeshiva University, New York, New York 10016, USA}
\affiliation{$^2$Instituto de F{\'i}sica, Universidad Aut\'onoma de Puebla, Apartado Postal J-48, Puebla, Puebla, 72570, Mexico}

\date{\today}

\begin{abstract}
Despite being ubiquitous, out-of-equilibrium quantum systems are much less understood than systems at equilibrium. Progress in the field has benefited from a symbiotic relationship between theoretical studies and new experiments on coherent dynamics. The present work strengthens this connection by providing a general picture of the relaxation process of isolated lattice many-body quantum systems that are routinely studied in experiments  with cold atoms, ions traps, and nuclear magnetic resonance. We show numerically and analytically that the long-time decay of the probability for finding the system in its initial state necessarily shows a power-law behavior $\propto  t^{ - \gamma }$. This happens independently of the details of the system, such as integrability, level repulsion, and the presence or absence of disorder. Information about the spectrum, the structure of the initial state, and the number of particles that interact simultaneously is contained in the value of  $\gamma$. From it, we can anticipate whether the initial state will or will not thermalize. 
\end{abstract}


\maketitle

{\em Introduction.} A great deal of effort has recently been put into improving our understanding of isolated many-body quantum systems quenched far from equilibrium.  This is in part motivated by the possibility of investigating the coherent evolution of these systems for long times with different experimental setups,  including those with ultracold atoms~\cite{Trotzky2012}, trapped ions~\cite{Jurcevic2014,Richerme2014}, and nuclear magnetic resonance~\cite{Cappellaro2007,Kaur2013}. Aligned with these efforts, this work characterizes and justifies the dynamical behavior at different time scales of experimentally accessible integrable and chaotic lattice many-body quantum systems with and without disorder. From this analysis, a new criterion, based exclusively on dynamics, is introduced for identifying which systems can thermalize.

The survival probability (probability for finding the system in its initial state at time $t$) and the Loschmidt echo (measure of the revival of the initial state after a time-reversal operation) have been extensively considered in the analysis of out-of-equilibrium quantum systems~\cite{MugaBook,Campo2011,Campo2016,Ketzmerick1992,Huckestein1994,Huckestein1999,Gorin2006,Goussev2012}. Several works tried to establish a correspondence between the initial exponential or Gaussian decays with quantum chaos~\cite{Gorin2006,Goussev2012,Flambaum2001a,Izrailev2006,KotaBook,Haldar2016} and others focused on the onset of power-law decays at long times~\cite{MugaBook,Campo2011,Campo2016,Ketzmerick1992,Huckestein1994,Huckestein1999}. In the case of continuous models, the algebraic behavior of the survival probability has been associated with the presence of bounds in the spectrum~\cite{MugaBook,Campo2011,Campo2016} while in disordered noninteracting systems at the metal-insulator transition,  the power-law exponent has been related with fractal dimensions~\cite{Ketzmerick1992,Huckestein1994,Huckestein1999}. Exchanges between these different communities have been very limited. Here, we unify these multiple perspectives into a single framework and use it to describe the evolution of the survival probability of lattice many-body quantum systems.

The survival probability (or fidelity) of the initial state is defined as 
\begin{eqnarray}
F(t) &\equiv& \left| \langle \Psi(0) | e^{-i H t} | \Psi(0) \rangle \right|^2  
\nonumber \\
&=& \left|\sum_{\alpha} |C_{\alpha}^{(0)} |^2 e^{-i E_{\alpha} t}  \right|^2 = 
\left| \int \!\! dE e^{-i E t} \rho_0(E) \right|^2 
\label{eq:fidelity}
\end{eqnarray}
where $E_{\alpha} $ are the eigenvalues of the system Hamiltonian $H$, $C_{\alpha}^{(0)}\!=\!\langle \psi_{\alpha} | \Psi(0) \rangle$ are the overlaps of the initial state $| \Psi(0) \rangle$ with the eigenstates $|\psi_{\alpha} \rangle $ of $H$, and 
$\rho _0(E) = \sum_\alpha  |C_\alpha ^{(0)} |^2 \delta (E - E_\alpha ) $
is the energy distribution of $| \Psi(0) \rangle$ weighted by the components  $|C_{\alpha} ^{(0)} |^2$, the so-called local density of states (LDOS). The survival probability is the absolute square of the Fourier transform of the LDOS. All information about the evolution of $F(t)$ is contained in $\rho_0(E)$.

We verified that the initial decay of the survival probability is dissociated from the regime (integrable or chaotic) of the Hamiltonian~\cite{Torres2014PRA,Torres2014NJP,Torres2014PRE,Torres2014PRAb,Torres2015,Torres2015BJP}, but depends on the strength of the perturbation.  We now show that at long times, regardless of how fast the initial evolution may be, the dynamics necessarily slows down and becomes power-law, $F(t) \propto t^{-\gamma}$. The characterization of the long-time dynamics and its connection with the viability of thermalization are the central topics of this work. 

We show that in realistic lattice many-body quantum systems with two-body interactions, $0 \leq \gamma \leq 2$. The value of the power-law exponent indicates the level of delocalization of the initial state in the energy eigenbasis. When the initial state samples only a portion of the Hilbert space and the LDOS is sparse, $\gamma<1$ and thermalization is not expected. When the initial state is chaotic, so that its components $C_\alpha ^{(0)}$ are uncorrelated and spread over its entire energy shell~\cite{Santos2012PRE,Santos2012PRL,Casati1996,Borgonovi2016}, thermalization should occur~\cite{ZelevinskyRep1996,Borgonovi2016,Santos2011PRL,Torres2013,TorresKollmar2015,He2012,Rigol2015}. In particular, when the LDOS is ergodically filled, then $\gamma=2$. From the values of $\gamma$, one can thus anticipate whether the initial state will or will not thermalize. We also discuss the non-realistic scenario of full random matrices, where the power-law exponent reaches the upper bound $\gamma=3$.

{\em Time scales.}   The system is initially prepared in an eigenstate of the unperturbed Hamiltonian $H_0$, which is abruptly quenched into $H=H_0 + g V$, where $g$ is the strength of the perturbation $V$. At very short times, the decay of the survival probability is quadratic~\cite{Wilkinson1997}, as derived from the expansion $F(t \ll \sigma_{0}^{-1})  \approx 1-\sigma_{0}^2 t^2$, where $\sigma_0=[\sum_{\alpha} |C_{\alpha}^{(0)} |^2 (E_{\alpha} - E_0)^2]^{1/2}$ is the width of the LDOS and $E_0= \sum_{\alpha} |C_{\alpha}^{(0)}|^2 E_{\alpha}=  \langle \Psi(0) |H|  \Psi(0) \rangle$ is the energy of the initial state. 

After the initial quadratic behavior, whether $F(t)$ switches or not to an exponential decay depends on the strength $g$ of the perturbation. The exponential decay is valid in the Fermi golden rule regime, where the typical matrix elements of $gV$ are larger than the mean level spacing and the LDOS has a Lorentzian form. However, for very strong perturbations, $g\rightarrow 1$, the LDOS is broader. In many-body quantum systems with two-body interactions, where the density of states is Gaussian~\cite{Brody1981,Kota2001,Zangara2013}, the limiting shape of the LDOS is also Gaussian, resulting in the Gaussian decay $F(t)= \exp(-\sigma_0^2 t^2)$ \cite{Torres2014PRA,Torres2014NJP,Torres2014PRE,Torres2014PRAb,Torres2015,Torres2015BJP,Flambaum2001a,Izrailev2006,noteFast}. 
Exponential and Gaussian decays can thus occur in both integrable and chaotic models~\cite{Torres2014PRA,Torres2014NJP,Torres2014PRE,Torres2014PRAb,Torres2015,Torres2015BJP}. The picture becomes more subtle at long times, where the power-law behavior $\propto t^{-\gamma}$ emerges and the filling of the LDOS plays a key role.  

{\em Causes of the power-law decay.} We discuss two distinctive causes for the long-time algebraic decay of the survival probability. 

\underline{Case 1} is related to the unavoidable presence of a lower bound $E_{\rm low}$ in the energy spectrum of any real quantum system.  This point was put forward already in 1958 \cite{Khalfin1958} and in several other early works~\cite{Nussenzweig1961,Ersak1969,Fleming1973,Knight1977,Fonda1978,Sluis1991}. At long times, the energy bound leads to the partial reconstruction of the initial state. This results in the power-law decay of continuous many-particle models~\cite{MugaBook,Campo2011,Campo2016} and, as explained here, also of finite lattice many-body quantum systems with ergodically filled LDOS. 

\underline{Case 2} is induced by the correlations that are present in nonchaotic eigenstates. They are typical of disordered systems undergoing localization with~\cite{Torres2015,Torres2015BJP} or without interactions~\cite{Ketzmerick1992,Huckestein1994,Huckestein1999} and, as argued here, appear also in clean integrable systems. The power-law exponent due to correlations is smaller than that resulting from energy bounds.

The exponents of case 1 can be derived from asymptotic expansions of the integral form of Eq.~(\ref{eq:fidelity}), assuming that $\rho_0(E)$ is absolutely integrable~\cite{Fock1947} and that its derivatives exist and are continuous in $[E_{\rm low}, \infty]$. Two scenarios are identified~\cite{Erdelyi1956,Urbanowski2009}:

(i) If $\lim _{E\rightarrow E_{\rm low}} \rho_0(E)>0$, then at long times $F(t) \! \propto \! t^{-2}$. 

(ii) If $\rho_0(E)$ decays abruptly close to the lower bound, such that $\rho_0(E) = (E-E_{\rm low})^{\xi} \eta(E)$ with  $0<\xi<1$ and $\lim _{E\rightarrow E_{\rm low}} \eta(E)>0$, then $F(t) \propto  t^{-2(\xi+1)}$. 

These results have been obtained for continuous functions. Yet we show that they remain valid even in the case of discrete spectra provided $|\Psi(0)\rangle$ is chaotic and the LDOS is ergodically filled. 

To determine if the initial state is chaotic, one performs scaling analysis of the inverse participation ratio (IPR) of $|\Psi(0) \rangle$ written in the energy eigenbasis, $\text{IPR}_{0}\equiv \sum_{\alpha} |C_{\alpha}^{(0)}|^4$. $\text{IPR}_{0}^{-1}$ is the effective number of energy eigenstates contributing to the initial state. A chaotic $| \Psi(0) \rangle$ samples most energy eigenbasis without any bias, so $\text{IPR}_{0} \propto {\cal D}^{-1}$, where ${\cal D}$ is the dimension of the Hilbert space. Hence, as the system size increases, $\rho_0(E)$ becomes homogeneously filled and close to an absolutely integrable function. An illustrative example is that of an arbitrary initial state projected onto the eigenstates of a full random matrix (FRM). Since these eigenstates are pseudo-random vectors, the overlaps $C_{\alpha}^{(0)}$ are random variables and $\text{IPR}_{0}\sim3/{\cal D}$~\cite{ZelevinskyRep1996}. Even though {\em realistic} chaotic many-body quantum systems with two-body interactions are not described by FRMs, because their Hamiltonian matrices are banded, sparse, and random elements may not even exist, they still follow random matrix statistics and their bulk eigenstates are close to random vectors~\cite{ZelevinskyRep1996,Santos2010PRE,Santos2010PREb}. After a strong perturbation into such Hamiltonians,  initial states with energies away from the edges of the spectrum also give very filled LDOS~\cite{ZelevinskyRep1996,Santos2012PRE,Santos2012PRL,Borgonovi2016,Casati1996,Santos2011PRL,Torres2013,TorresKollmar2015,He2012}. 

Case 1(i) holds for realistic chaotic many-body quantum systems, where the LDOS is Gaussian, which leads to $\gamma=2$. For FRM, the LDOS is a semicircle~\cite{Torres2014PRA}, so case 1(ii) applies and $\gamma=3$.  

In case 2, the power-law exponent is obtained from the correlation function $C(\omega ) \equiv  \sum_{\alpha ,\beta } | C_{\beta}^{(0)} |^2 | C_{\alpha}^{(0)} |^2 \delta (E_{\alpha } \!-\! E_{\beta } \!-\! \omega )  $ present in $F(t)  = \int_{ - \infty }^\infty  d\omega e^{i\omega t} C(\omega ) $. A power-law decay of $C(\omega \rightarrow 0 ) \propto \omega^{\gamma -1} $, with $\gamma<1$, leads to $F(t) \propto t^{-\gamma}$ \cite{Chalker1988,Chalker1990,Ketzmerick1992,Huckestein1994,Huckestein1999,Kravtsov2011}. The more correlated the components of $|\Psi(0) \rangle$, the smaller the exponent $\gamma$. This exponent coincides with the fractal dimension $\phi$ obtained from the scaling analysis of $\text{IPR}_{0} \propto {\cal D}^{-\phi}$. This relation was found in studies of Anderson localization~\cite{Ketzmerick1992,Huckestein1994,Huckestein1999} and of many-body localization~\cite{Torres2015,Torres2015BJP}. We show that it holds also in noninteracting integrable models.

This work analyzes how $\gamma$ depends on the properties of the spectrum, the structure of the initial state, and the number of particles that interact simultaneously. We consider finite many-body quantum systems described by realistic lattice models with two-body interactions and by banded random matrices. All accessible power-law exponents are reached with the disordered models, while with the clean Hamiltonians, we study some specific values.

{\em Realistic many-body quantum systems.} We consider one-dimensional spin-1/2 models with $L$ sites described by the following Hamiltonian,
\begin{eqnarray}
&&H = \sum_{n=1}^L h_n  S_n^z+ H_{NN} + \lambda H_{NNN},
\label{ham} \\
&& H_{NN} = J \sum_{n} \left(S_n^x S_{n+1}^x + S_n^y S_{n+1}^y +\Delta S_n^z S_{n+1}^z \right) \;,
\nonumber \\
&& H_{NNN} = \sum_{n} J \left(S_n^x S_{n+2}^x + S_n^y S_{n+2}^y +\Delta S_n^z S_{n+2}^z \right) \;.
\nonumber 
\end{eqnarray}
It contains nearest-neighbor (NN) and possibly also next-nearest-neighbor (NNN) couplings; $\hbar=1$ and $S^{x,y,z}_n$ are the spin operators on site $n$. $h_n$ are random numbers from a uniform distribution $[-h,h]$; the system is clean when $h=0$ and disordered otherwise. $J$ is the coupling strength, $\Delta$ the anisotropy parameter, and $\lambda$ the ratio  between NNN and NN couplings. $J=1$ sets the energy scale. The Hamiltonian conserves total spin in the $z$ direction ${\cal S}^z$. We work with the largest subspace ${\cal S}^z=0$ of dimension ${\cal D}=L!/(L/2)!^2$.

The integrable limits of $H$ include the clean noninteracting $XX$ ($\Delta, \lambda, h= 0$) and the clean interacting $XXZ$ ($\Delta \neq 0$, $\lambda, h=0$) models. The system becomes chaotic as $\lambda $ increases from zero~\cite{Hsu1993,Kudo2005,Santos2009JMP,Gubin2012} and the level spacing distribution changes from Poisson~\cite{noteXX} to a Wigner-Dyson form~\cite{Guhr1998}. It also becomes chaotic when the disorder strength increases from zero and $h<J$ \cite{Avishai2002,Santos2004,Dukesz2009}. 

The initial states considered are site-basis vectors, where the spin on each site either points up or down in the $z$-direction. An example is the  experimentally~\cite{Trotzky2008} accessible N\'eel state, $ |\uparrow \downarrow \uparrow \downarrow \uparrow \downarrow \uparrow \downarrow   \ldots \rangle $, that has been extensively used in studies of the dynamics of integrable spin systems. Site-basis vectors evolve under $H$ (\ref{ham}) after a strong perturbation, where the anisotropy parameter is quenched from $\Delta \rightarrow \infty$ to a finite value. The envelope of the LDOS for these initial states is therefore Gaussian.

\begin{figure}[htb]
\centering
\includegraphics*[width=3.1in]{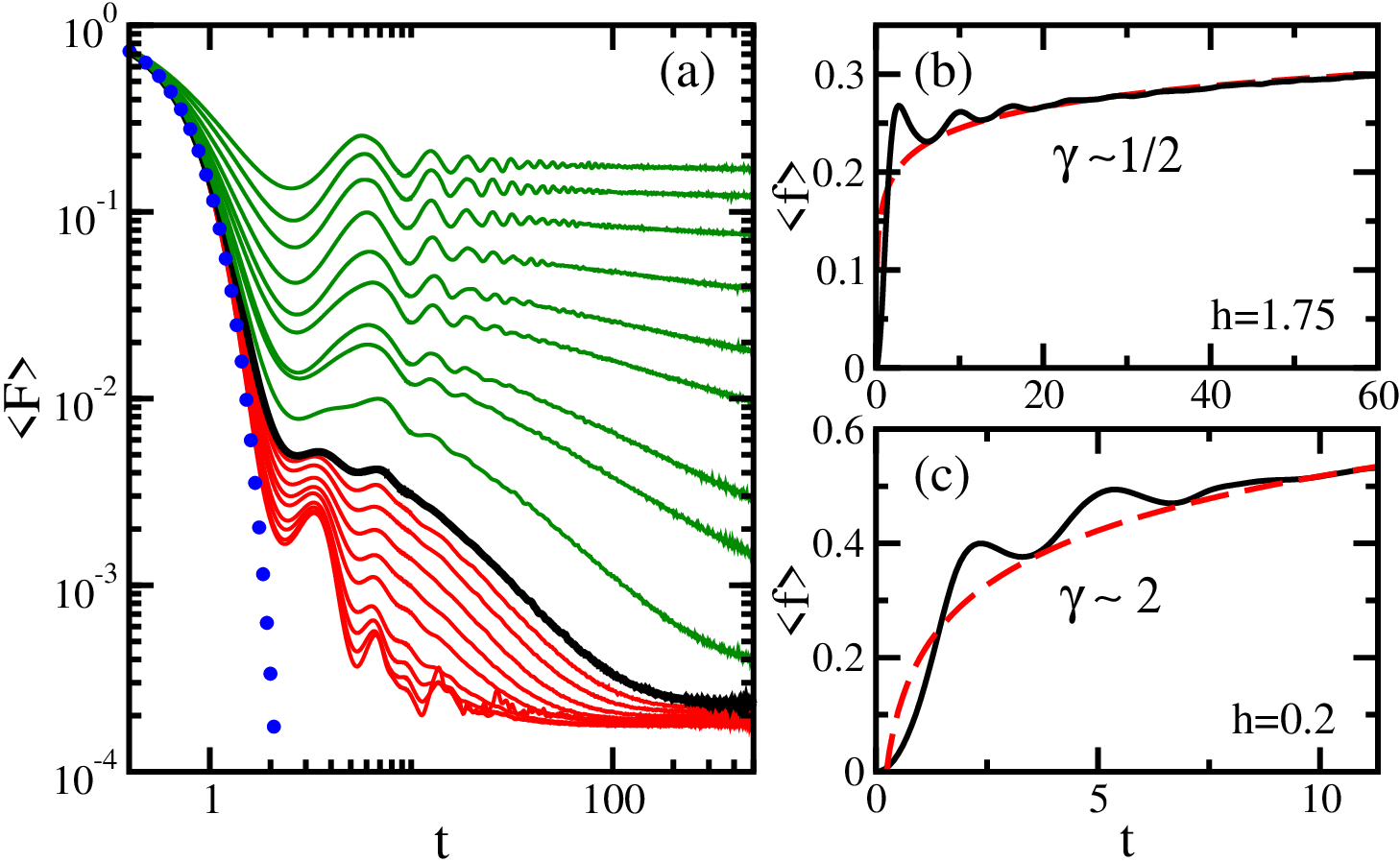}
\caption{Survival probability (a) and $f(t)$ (b), (c). In (a): from bottom to top, $h=0.2, 0.3, \ldots 0.9$, $h=0.95, 1, 1.25, \ldots 3$, $h=3.5$. Thick solid line: $h=1$ with $\gamma \sim 1$. Circles: analytical Gaussian decay $F(t)= \exp(-\sigma_0^2 t^2)$. In (b) and (c): Numerical curve (solid), ${\rm const}\!-\!L^{-1} \!\ln t^{-\gamma}$ (dashed). Averages over $10^5$ data of disorder realizations and initial states with $E_0 \sim 0$; $L=16$, closed boundaries.}
\label{fig:disorder}  
\end{figure}

{\em Realistic disordered systems}. Figure~\ref{fig:disorder} shows the survival probability of site-basis vectors evolving under $H$ (\ref{ham}) with $\Delta=1$, $\lambda=0$ and various values of $h$. The initial decay is Gaussian, as expected from the Gaussian LDOS. It agrees very well with the analytical expression $F(t)= \exp(-\sigma_0^2 t^2)$, as seen for the bottom curve of Fig.~\ref{fig:disorder} (a). Subsequently the dynamics slows down and becomes a power-law for all curves. 

When the disorder strength is small, $0<h<1$, the system is chaotic and the LDOS is very filled. This is corroborated from the analysis of level statistics and by computing the inverse participation ratio averaged over initial states and random realizations. One finds that $\langle \text{IPR}_{0} \rangle \propto {\cal D}^{-1}$. For the value of $h$ where $\langle \text{IPR}_{0}\rangle $ is maximum, the decay of $F(t)$ at long times is $\propto t^{-2}$, as illustrated with the bottom curve in Fig.~\ref{fig:disorder} (a). For other values of $h$ in $(0,1]$, we have the intermediate region, where $1\leq \gamma <2$.  These values may result from a competition between weak correlations and energy bounds, but this needs to be further investigated.

The bottom curve of Fig.~\ref{fig:disorder}(a) is isolated in Fig.~\ref{fig:disorder}(c), which shows the rescaled survival probability $f(t) = - (1/L) \ln F(t)$ \cite{Heyl2013,Andraschko2014}. For $L\gg 1$, this quantity is independent of $L$ \cite{Touchette2009}.  Figure~\ref{fig:disorder} (c) is a clear example of the power-law decay caused by energy bounds [case 1(i)]. The Fourier transform of a Gaussian LDOS that has lower $E_{\rm low}$ and upper $E_{\rm up}$ bounds, as in our case, leads to 
$ F(t) =  \frac{e^{ - \sigma _0^2 t^2} }{4{\cal N}^2}  \left|   \text{erf} \left( \frac{E_0 - E_{\rm low} + i\sigma _0^2t}{\sqrt{2} \sigma _0 } \right)   
 -  \text{erf} \left( \frac{E_0 - E_{\rm up} + i\sigma _0^2t}{\sqrt{2} \sigma _0 } \right)  \right|^2 $, where $\text{erf}$ is the error function and ${\cal N}$ is a normalization constant that depends on $L$ through the energy bounds and $\sigma _0$. At long times, after dropping the oscillations from the sinusoidal term $\cos [t (E_{\rm up}+E_{\rm low})]$,  the expression becomes
$F(t\gg \sigma_0^{-1}) \simeq (2\pi\sigma_0^2t^2{\cal N}^2)^{-1} \sum_{k=\rm up,\rm low} e^{-(E_k-E_0)^2/\sigma_0^2}$,
from where the $t^{-2}$ power-law decay is evident.

When $h=1$, we get $\gamma \sim 1$. This curve is depicted with a thick line in Fig.~\ref{fig:disorder} (a). Above this line, $h>1$ and $\gamma<1$. An example with $\gamma \sim 1/2$ is isolated in Fig.~\ref{fig:disorder} (b). This $\gamma$ is close to the exponent $\phi$ obtained from the scaling analysis of $\text{IPR}_0 \propto {\cal D}^{-\phi}$ \cite{Torres2015}. This example belongs to case 2. 

Figure~\ref{fig:disorder} (a) demonstrates that with the disordered $XXZ$ model, we can obtain all power-law exponents accessible to realistic lattice many-body quantum systems with two-body interactions. By varying $h$, every $\gamma \in [0,2]$ can be reached.

{\em Banded random matrices}. Algebraic decays faster than $t^{-2}$ also signal the ergodic filling of the LDOS. They are possible if instead of two-body interactions, many-body random interactions are included. As the number of particles that interact simultaneously grows, increasing the number of uncorrelated nonzero elements in the Hamiltonian matrix, the density of states transitions from Gaussian to a semicircle~\cite{Brody1981}. The latter is typical of FRMs \cite{Guhr1998}. This transition is reflected also in the shape of the LDOS \cite{Torres2014PRA,Torres2014NJP,Wigner1955,Casati1996,Fyodorov1996}. The Fourier transform of a semicircle gives $F(t) =[{\cal J}_1( 2 \sigma_{0} t)]^2/(\sigma_{0}^2 t^2)$, where ${\cal J}_1$ is the Bessel function of the first kind~\cite{Torres2014PRA,Torres2014NJP}. The decay at short times is faster than Gaussian and the asymptotic expansion reveals a power-law decay with $\gamma =3$, $F(t\gg \sigma_{0}^{-1}) \simeq [1 - \sin(4 \sigma_{0} t)]/(2 \pi \sigma_{0}^3 t^3)$. This is an example of case 1(ii), where for the semicircle, $\xi=1/2$, $\eta(E) = (2\pi \sigma_0^2)^{-1} (2\sigma_0 - E)^{1/2}$, and $E_{\rm low}=-2\sigma_0$.
\\
\begin{figure}[htb]
\centering
\includegraphics*[width=2.7in]{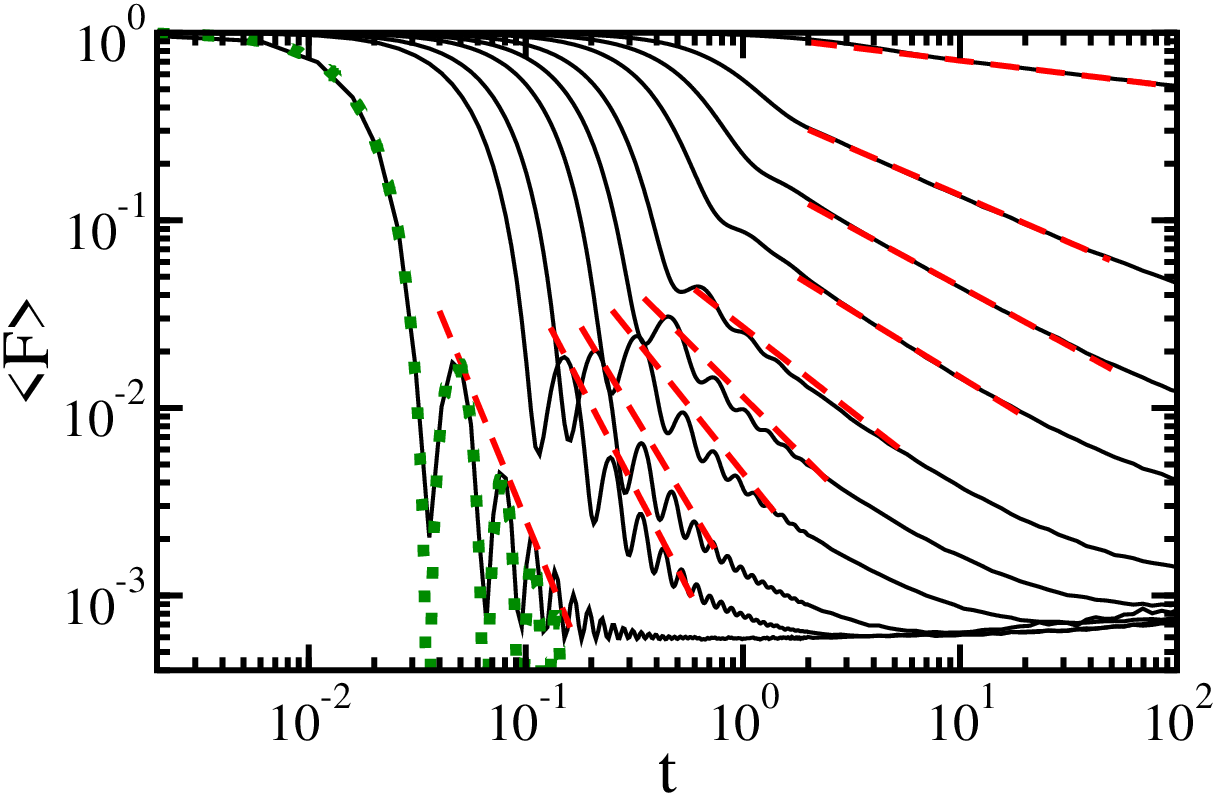}
\caption{Survival probability for basis vectors evolving under PBRM with $b=0.1,0.5,1,2,5,10,20,50,100,3000 $ (solid) from top to bottom. They correspond respectively to the fitted $\gamma \sim 0.1, 0.5, 0.6, 0.7, 0.9, 1.2, 1.4, 1.9, 2.2, 2.8$ (dashed). Analytical $F(t) =[{\cal J}_1( 2 \sigma_{0} t)]^2/(\sigma_{0}^2 t^2)$ (dotted); ${\cal D}=3432$. Averages over 100 realizations and 343 initial states with $E_0 \sim 0$.
}
\label{fig:pbrm}  
\end{figure}

To illustrate the increase of the value of $\gamma$ from $2$ to the upper bound $\gamma=3$, we consider power-law banded random matrices (PBRM) \cite{Mirlin1996,Mirlin2000,Evers2008}. Despite the success of FRMs in describing statistically the spectrum of complex systems, they imply the unphysical scenario of all particles interacting simultaneously. Banded random matrices were introduced~\cite{Wigner1955} in an effort to better approach random matrices to real systems. We use PBRMs that preserve time reversal symmetry and whose elements are real random numbers from a Gaussian distribution~\cite{Varga2002}: $\langle H_{nn}\rangle \!=\! 0 $, $\langle H_{nn}^2 \rangle \!=\!2 $, $\langle H_{nm}^2 \rangle = 1/ [1+|(n-m)/b|^2] $ for $n\neq m$. The value of $b$ determines how fast the elements decrease as they move away from the diagonal. When $b\rightarrow {\cal D}$, the PBRM coincides with a FRM.

In Fig.~\ref{fig:pbrm}, we show the survival probability for PBRMs with different values of $b$. As $b$ grows from $\sim50$ to ${\cal D}$ and the LDOS transitions from case 1(i) to case 1(ii), $\gamma$ increases from 2 to 3. In the other direction, as $b$ decreases below $50$, the eigenstates become less spread out and $\gamma$ decreases below $2$.  With PBRMs, we obtain a general picture of the behavior of the survival probability, covering all values of $\gamma$, without any restriction to a specific model.

\begin{figure}[htb]
\centering
\includegraphics*[width=3.1in]{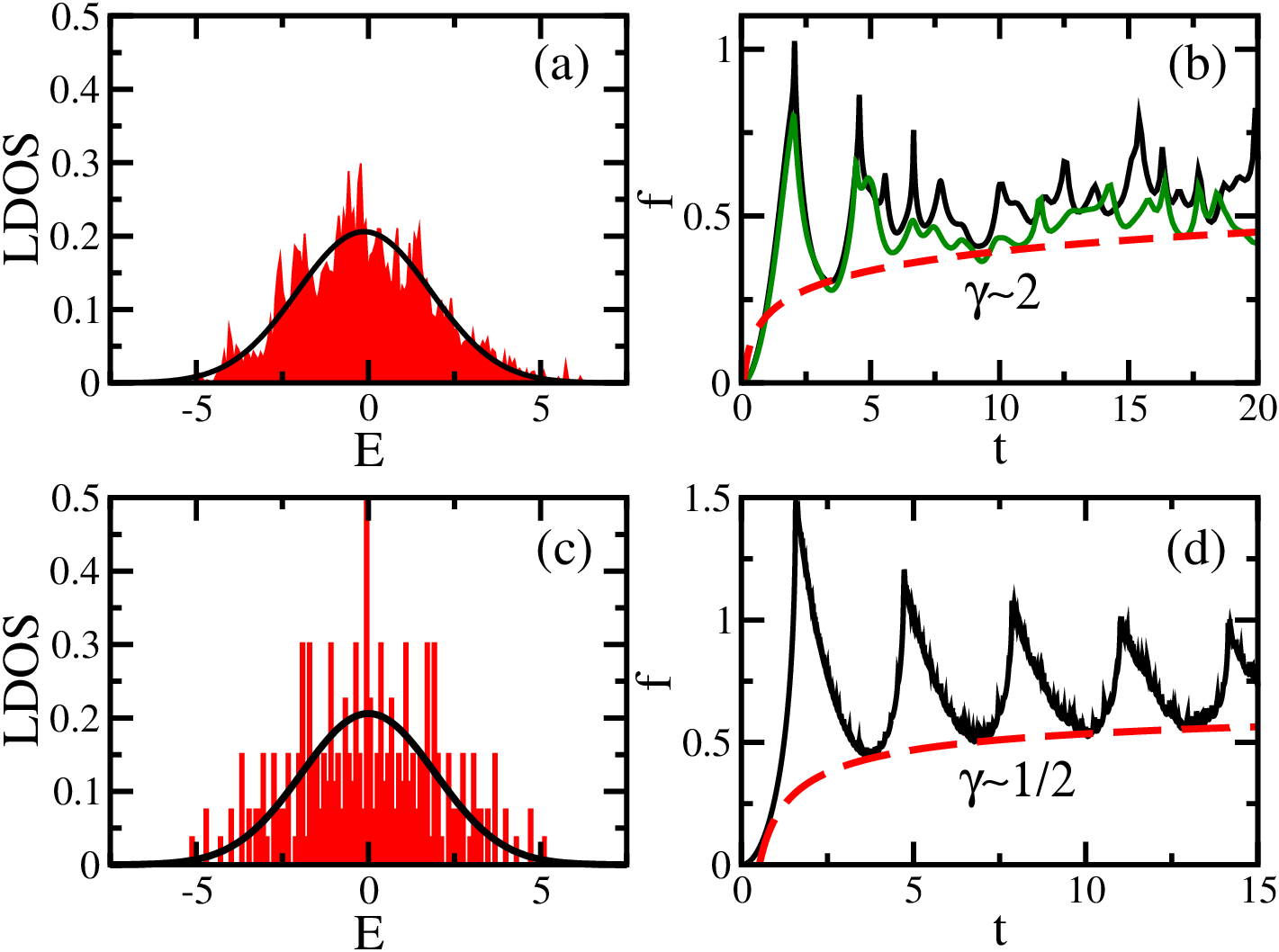}
\caption{LDOS [(a),(c)] and $f(t)$ [(b),(d)] for the N\'eel state under the chaotic open $H$ (\ref{ham}) with $h=0,\Delta=1/2,\lambda=1$ [(a),(b)] and under the closed $XX$ $H$ [(c),(d)]. 
(a),(c): Numerical LDOS (shaded area) and Gaussian envelope (solid line). (b): Numerical results for $L=22$ (light), $L=24$ (dark), and ${\rm const}\!-\!L^{-1} \!\ln t^{-2}$ (dashed). (d): $L=400$ (solid) and $f_{XX}^{\text{N\'eel}}(t)$ (dashed).}
\label{fig:chaos}  
\end{figure}

{\em Realistic clean systems}. In Fig.~\ref{fig:chaos}, we study the N\'eel state evolving under a clean chaotic Hamiltonian [Figs. 3(a) and 3(b)] and under the $XX$ Hamiltonian [Figs. 3(c) and 3(d)]. The envelope of the LDOS is Gaussian in both cases (a) and (c), but visibly sparse in Fig.~\ref{fig:chaos}(c).

In Fig.~\ref{fig:chaos}(b),  we observe a power-law decay $\propto t^{-2}$. The agreement between the $t^{-2}$ decay (dashed line) and our numerical results (solid lines) suggests that the LDOS must be ergodically filled and that thermalization should occur. Indeed, the inverse participation ratio of the N\'eel state in Fig.~\ref{fig:chaos}  (a) gives $\text{IPR}_{0} \propto {\cal D}^{-1}$ and several studies for this model confirm thermalization~\cite{rigol09STATa,rigol09STATb,Santos2011PRL,Torres2013}.  We found $\gamma=2$ also for periodic boundary conditions; chaotic models with different values of $\lambda$ and $\Delta$, including $\Delta=0$; and other initial states.  A $t^{-2}$ decay has also been speculated for the chaotic Ising model with longitudinal and transverse fields~\cite{Monnai2014}.


An analytical expression exists for $F(t)$ for the N\'eel state evolving under the periodic $XX$ model~\cite{Andraschko2014,Mazza2016}.  Its expansion for long times, $L t^{-1/2} \rightarrow 0$, gives
$f_{XX}^{\text{N\'eel}}(t) \to  - L^{-1} \ln \left[ 2^{ - L} \left( 1 + 2^{-1} L t^{-1/2} \right) \right]$, as indeed confirmed with the dashed line in Fig.~\ref{fig:chaos}(d).  Such small $\gamma$ indicates that the LDOS is not ergodically filled, as seen in Fig.~\ref{fig:chaos}(c) and corroborated below by calculating $\text{IPR}_0$. 

Among the total ${\cal D} =L!/(L/2){!^2}$ components of the N\'eel state, only $2^{L/2}$ are nonzero and they are all equal,  $|C_{\alpha}^{(0)}|^2 = 2^{-L/2} $ \cite{Mazza2016}. This means that $\text{IPR}_0  = 2^{ - L/2}$. Using the Stirling approximation for large $L$, we have that $\ln {\cal D} \simeq L \ln 2$. From $\ln \text{IPR}_{0} $ vs $\ln {\cal D}$, we find that $\text{IPR}_0 \simeq {\cal D}^{-1/2}$, so $\phi=1/2$. One sees that, similarly to what is done in disordered systems~\cite{Torres2015,Torres2015BJP}, the power-law exponent for the N\'eel state in the $XX$ model, $\gamma=1/2$, can also be extracted from the scaling analysis of $\text{IPR}_0$.

The nonzero $|C_{\alpha}^{(0)}|^2$ are spread out in energy, resulting in a very sparse and inhomogeneous LDOS. The nonergodicity of this state indicates that thermalization should not occur. One way to confirm thermalization is by verifying the coincidence of the diagonal entropy $S_{\rm d}\!=\!-\!\sum_{\alpha}|C_{\alpha}^{(0)}|^2 \ln |C_{\alpha}^{(0)}|^2$ \cite{Polkovnikov2011} and  the thermodynamic entropy, $S_{\rm th}=\ln \sum_{\alpha} e^{-E_{\alpha}/T} -  (\sum_{\alpha} E_{\alpha} e^{-E_{\alpha}/T})/(T \sum_{\alpha} e^{-E_{\alpha}/T} )$ \cite{Santos2011PRL}. Here, $S_{\rm d}=(L/2)\ln 2$ and  $S_{\rm th}=\ln {\cal D}$ [Note that the N\'eel state has $E_0=0$ and thus infinite temperature $T$]. The two entropies do not coincide even in the thermodynamic limit, where $(S_{\rm th} - S_{\rm d})/L= \ln \sqrt{2}$.

{\em  Conclusions.} We have shown that the long-time decay of the survival probability in isolated lattice many-body quantum systems is algebraic, $F(t)\propto t^{-\gamma}$, be the system integrable or chaotic, interacting or noninteracting, clean or disordered. The entire range of $\gamma \in [0,3]$ can be reached with banded random matrices, while for realistic systems with two-body interactions, $\gamma \in [0,2]$. From the value of $\gamma$, we infer how much delocalized the initial state is in the energy eigenbasis. This provides a way to identify whether the initial state will thermalize based exclusively on its dynamics. Exponents $\gamma \geq 2$  signal ergodicity and therefore thermalization. Advantages of this approach to the problem of thermalization include the following: any initial state can be considered, numerical methods other than exact diagonalization are available for analyzing dynamics, and a natural connection is established with experiments that routinely study the dynamics of many-body quantum systems.


{\em Acknowledgments.} This work was supported by the NSF Grant No.~DMR-1147430 and  Yeshiva University. EJTH acknowledges funding from CONACyT, PRODEP-SEP and VIEP-BUAP, Mexico. We thank Adolfo del Campo, Yevgeny Bar Lev, and Marcos Rigol for useful discussions.


\begin{thebibliography}{78}
\expandafter\ifx\csname natexlab\endcsname\relax\def\natexlab#1{#1}\fi
\expandafter\ifx\csname bibnamefont\endcsname\relax
  \def\bibnamefont#1{#1}\fi
\expandafter\ifx\csname bibfnamefont\endcsname\relax
  \def\bibfnamefont#1{#1}\fi
\expandafter\ifx\csname citenamefont\endcsname\relax
  \def\citenamefont#1{#1}\fi
\expandafter\ifx\csname url\endcsname\relax
  \def\url#1{\texttt{#1}}\fi
\expandafter\ifx\csname urlprefix\endcsname\relax\def\urlprefix{URL }\fi
\providecommand{\bibinfo}[2]{#2}
\providecommand{\eprint}[2][]{\url{#2}}

\bibitem[{\citenamefont{Trotzky {\it et~al.}}(2012)\citenamefont{Trotzky, Chen,
  Flesch, McCulloch, Schollw\"ock, Eisert, and Bloch}}]{Trotzky2012}
\bibinfo{author}{\bibfnamefont{S.}~\bibnamefont{Trotzky}},
  \bibinfo{author}{\bibfnamefont{Y.-A.} \bibnamefont{Chen}},
  \bibinfo{author}{\bibfnamefont{A.}~\bibnamefont{Flesch}},
  \bibinfo{author}{\bibfnamefont{I.~P.} \bibnamefont{McCulloch}},
  \bibinfo{author}{\bibfnamefont{U.}~\bibnamefont{Schollw\"ock}},
  \bibinfo{author}{\bibfnamefont{J.}~\bibnamefont{Eisert}}, \bibnamefont{and}
  \bibinfo{author}{\bibfnamefont{I.}~\bibnamefont{Bloch}},
  \bibinfo{journal}{Nature Phys.} \textbf{\bibinfo{volume}{8}},
  \bibinfo{pages}{325} (\bibinfo{year}{2012}); \bibinfo{author}{\bibfnamefont{T.}~\bibnamefont{Fukuhara}}
  \bibnamefont{et~al.}, \bibinfo{journal}{Nat. Phys.}
  \textbf{\bibinfo{volume}{9}}, \bibinfo{pages}{235} (\bibinfo{year}{2013}).

\bibitem[{\citenamefont{Jurcevic et~al.}(2014)\citenamefont{Jurcevic, Lanyon,
  Hauke, Hempel, Zoller, Blatt, and Roos}}]{Jurcevic2014}
\bibinfo{author}{\bibfnamefont{P.}~\bibnamefont{Jurcevic}},
  \bibinfo{author}{\bibfnamefont{B.~P.} \bibnamefont{Lanyon}},
  \bibinfo{author}{\bibfnamefont{P.}~\bibnamefont{Hauke}},
  \bibinfo{author}{\bibfnamefont{C.}~\bibnamefont{Hempel}},
  \bibinfo{author}{\bibfnamefont{P.}~\bibnamefont{Zoller}},
  \bibinfo{author}{\bibfnamefont{R.}~\bibnamefont{Blatt}}, \bibnamefont{and}
  \bibinfo{author}{\bibfnamefont{C.~F.} \bibnamefont{Roos}},
  \bibinfo{journal}{Nature (London)} \textbf{\bibinfo{volume}{511}},
  \bibinfo{pages}{202} (\bibinfo{year}{2014}).

\bibitem[{\citenamefont{Richerme et~al.}(2014)\citenamefont{Richerme, Gong,
  Lee, Senko, Smith, Foss-Feig, Michalakis, Gorshkov, and
  Monroe}}]{Richerme2014}
\bibinfo{author}{\bibfnamefont{P.}~\bibnamefont{Richerme}},
  \bibinfo{author}{\bibfnamefont{Z.-X.} \bibnamefont{Gong}},
  \bibinfo{author}{\bibfnamefont{A.}~\bibnamefont{Lee}},
  \bibinfo{author}{\bibfnamefont{C.}~\bibnamefont{Senko}},
  \bibinfo{author}{\bibfnamefont{J.}~\bibnamefont{Smith}},
  \bibinfo{author}{\bibfnamefont{M.}~\bibnamefont{Foss-Feig}},
  \bibinfo{author}{\bibfnamefont{S.}~\bibnamefont{Michalakis}},
  \bibinfo{author}{\bibfnamefont{A.~V.} \bibnamefont{Gorshkov}},
  \bibnamefont{and} \bibinfo{author}{\bibfnamefont{C.}~\bibnamefont{Monroe}},
  \bibinfo{journal}{Nature (London)} \textbf{\bibinfo{volume}{511}},
  \bibinfo{pages}{198} (\bibinfo{year}{2014}).

\bibitem[{\citenamefont{Cappellaro et~al.}(2007)\citenamefont{Cappellaro,
  Ramanathan, and Cory}}]{Cappellaro2007}
\bibinfo{author}{\bibfnamefont{P.}~\bibnamefont{Cappellaro}},
  \bibinfo{author}{\bibfnamefont{C.}~\bibnamefont{Ramanathan}},
  \bibnamefont{and} \bibinfo{author}{\bibfnamefont{D.~G.} \bibnamefont{Cory}},
  \bibinfo{journal}{Phys. Rev. Lett.} \textbf{\bibinfo{volume}{99}},
  \bibinfo{pages}{250506} (\bibinfo{year}{2007}).

\bibitem[{\citenamefont{Kaur et~al.}(2013)\citenamefont{Kaur, Ajoy, and
  Cappellaro}}]{Kaur2013}
\bibinfo{author}{\bibfnamefont{G.}~\bibnamefont{Kaur}},
  \bibinfo{author}{\bibfnamefont{A.}~\bibnamefont{Ajoy}}, \bibnamefont{and}
  \bibinfo{author}{\bibfnamefont{P.}~\bibnamefont{Cappellaro}},
  \bibinfo{journal}{New J. Phys.} \textbf{\bibinfo{volume}{15}},
  \bibinfo{pages}{093035} (\bibinfo{year}{2013}).

\bibitem[{\citenamefont{Muga et~al.}(2009)\citenamefont{Muga, Ruschhaupt, and
  del Campo}}]{MugaBook}
\bibinfo{author}{\bibfnamefont{J.~G.} \bibnamefont{Muga}},
  \bibinfo{author}{\bibfnamefont{A.}~\bibnamefont{Ruschhaupt}},
  \bibnamefont{and} \bibinfo{author}{\bibfnamefont{A.}~\bibnamefont{del
  Campo}}, \emph{\bibinfo{title}{Time in Quantum Mechanics}}
  (\bibinfo{publisher}{Springer}, \bibinfo{address}{London},
  \bibinfo{year}{2009}),Vol.2.

\bibitem[{\citenamefont{del Campo}(2011)}]{Campo2011}
\bibinfo{author}{\bibfnamefont{A.}~\bibnamefont{del Campo}},
  \bibinfo{journal}{Phys. Rev. A} \textbf{\bibinfo{volume}{84}},
  \bibinfo{pages}{012113} (\bibinfo{year}{2011}).

\bibitem[{\citenamefont{del Campo}(2016)}]{Campo2016}
\bibinfo{author}{\bibfnamefont{A.}~\bibnamefont{del Campo}},
  \bibinfo{journal}{New J. Phy.} \textbf{\bibinfo{volume}{18}},
  \bibinfo{pages}{015014} (\bibinfo{year}{2016}).

\bibitem[{\citenamefont{Ketzmerick et~al.}(1992)\citenamefont{Ketzmerick,
  Petschel, and Geisel}}]{Ketzmerick1992}
\bibinfo{author}{\bibfnamefont{R.}~\bibnamefont{Ketzmerick}},
  \bibinfo{author}{\bibfnamefont{G.}~\bibnamefont{Petschel}}, \bibnamefont{and}
  \bibinfo{author}{\bibfnamefont{T.}~\bibnamefont{Geisel}},
  \bibinfo{journal}{Phys. Rev. Lett.} \textbf{\bibinfo{volume}{69}},
  \bibinfo{pages}{695} (\bibinfo{year}{1992}).

\bibitem[{\citenamefont{Huckestein and Schweitzer}(1994)}]{Huckestein1994}
\bibinfo{author}{\bibfnamefont{B.}~\bibnamefont{Huckestein}} \bibnamefont{and}
  \bibinfo{author}{\bibfnamefont{L.}~\bibnamefont{Schweitzer}},
  \bibinfo{journal}{Phys. Rev. Lett.} \textbf{\bibinfo{volume}{72}},
  \bibinfo{pages}{713} (\bibinfo{year}{1994}).

\bibitem[{\citenamefont{Huckestein and Klesse}(1999)}]{Huckestein1999}
\bibinfo{author}{\bibfnamefont{B.}~\bibnamefont{Huckestein}} \bibnamefont{and}
  \bibinfo{author}{\bibfnamefont{R.}~\bibnamefont{Klesse}},
  \bibinfo{journal}{Phys. Rev. B} \textbf{\bibinfo{volume}{59}},
  \bibinfo{pages}{9714} (\bibinfo{year}{1999}).

\bibitem[{\citenamefont{Gorin et~al.}(2006)\citenamefont{Gorin, Prosen,
  Seligman, and \ifmmode \check{Z}\else \v{Z}\fi{}nidari\ifmmode~\check{c}\else
  \v{c}\fi{}}}]{Gorin2006}
\bibinfo{author}{\bibfnamefont{T.}~\bibnamefont{Gorin}},
  \bibinfo{author}{\bibfnamefont{T.}~\bibnamefont{Prosen}},
  \bibinfo{author}{\bibfnamefont{T.~H.} \bibnamefont{Seligman}},
  \bibnamefont{and} \bibinfo{author}{\bibfnamefont{M.}~\bibnamefont{\ifmmode
  \check{Z}\else \v{Z}\fi{}nidari\ifmmode~\check{c}\else \v{c}\fi{}}},
  \bibinfo{journal}{Phys. Rep.} \textbf{\bibinfo{volume}{435}},
  \bibinfo{pages}{33 } (\bibinfo{year}{2006}).

\bibitem[{\citenamefont{Goussev et~al.}(2012)\citenamefont{Goussev, Jalabert,
  Pastawski, and Wisniacki}}]{Goussev2012}
\bibinfo{author}{\bibfnamefont{A.}~\bibnamefont{Goussev}},
  \bibinfo{author}{\bibfnamefont{R.~A.} \bibnamefont{Jalabert}},
  \bibinfo{author}{\bibfnamefont{H.~M.} \bibnamefont{Pastawski}},
  \bibnamefont{and} \bibinfo{author}{\bibfnamefont{D.~A.}
  \bibnamefont{Wisniacki}}, \bibinfo{journal}{Scholarpedia}
  \textbf{\bibinfo{volume}{7}}, \bibinfo{pages}{11687} (\bibinfo{year}{2012}).

\bibitem[{\citenamefont{Flambaum and Izrailev}(2001)}]{Flambaum2001a}
\bibinfo{author}{\bibfnamefont{V.~V.} \bibnamefont{Flambaum}} \bibnamefont{and}
  \bibinfo{author}{\bibfnamefont{F.~M.} \bibnamefont{Izrailev}},
  \bibinfo{journal}{Phys. Rev. E} \textbf{\bibinfo{volume}{64}},
  \bibinfo{pages}{026124} (\bibinfo{year}{2001}).

\bibitem[{\citenamefont{Izrailev and
  Casta{\~{n}}eda-Mendoza}(2006)}]{Izrailev2006}
\bibinfo{author}{\bibfnamefont{F.~M.} \bibnamefont{Izrailev}} \bibnamefont{and}
  \bibinfo{author}{\bibfnamefont{A.}~\bibnamefont{Casta{\~{n}}eda-Mendoza}},
  \bibinfo{journal}{Phys. Lett. A} \textbf{\bibinfo{volume}{350}},
  \bibinfo{pages}{355} (\bibinfo{year}{2006}).

\bibitem[{\citenamefont{Kota}(2014)}]{KotaBook}
\bibinfo{author}{\bibfnamefont{V.~K.~B.} \bibnamefont{Kota}},
  \emph{\bibinfo{title}{Lecture Notes in Physics}}
  (\bibinfo{publisher}{Springer}, \bibinfo{address}{Heidelberg},
  \bibinfo{year}{2014}),Vol.884.

\bibitem[{\citenamefont{Haldar et~al.}(2016)\citenamefont{Haldar, Chavda, Vyas,
  and Kota}}]{Haldar2016}
\bibinfo{author}{\bibfnamefont{S.~K.} \bibnamefont{Haldar}},
  \bibinfo{author}{\bibfnamefont{N.~D.} \bibnamefont{Chavda}},
  \bibinfo{author}{\bibfnamefont{M.}~\bibnamefont{Vyas}}, \bibnamefont{and}
  \bibinfo{author}{\bibfnamefont{V.~K.~B.} \bibnamefont{Kota}},
  \bibinfo{journal}{J. Stat. Mech.} \textbf{\bibinfo{volume}{2016}},
  \bibinfo{pages}{043101} (\bibinfo{year}{2016}).

\bibitem[{\citenamefont{Torres-Herrera and
  Santos}(2014{\natexlab{a}})}]{Torres2014PRA}
\bibinfo{author}{\bibfnamefont{E.~J.} \bibnamefont{Torres-Herrera}}
  \bibnamefont{and} \bibinfo{author}{\bibfnamefont{L.~F.}
  \bibnamefont{Santos}}, \bibinfo{journal}{Phys. Rev. A}
  \textbf{\bibinfo{volume}{89}}, \bibinfo{pages}{043620}
  (\bibinfo{year}{2014}{\natexlab{a}}).

\bibitem[{\citenamefont{Torres-Herrera
  et~al.}(2014)\citenamefont{Torres-Herrera, Vyas, and Santos}}]{Torres2014NJP}
\bibinfo{author}{\bibfnamefont{E.~J.} \bibnamefont{Torres-Herrera}},
  \bibinfo{author}{\bibfnamefont{M.}~\bibnamefont{Vyas}}, \bibnamefont{and}
  \bibinfo{author}{\bibfnamefont{L.~F.} \bibnamefont{Santos}},
  \bibinfo{journal}{New J. Phys.} \textbf{\bibinfo{volume}{16}},
  \bibinfo{pages}{063010} (\bibinfo{year}{2014}).

\bibitem[{\citenamefont{Torres-Herrera and
  Santos}(2014{\natexlab{b}})}]{Torres2014PRE}
\bibinfo{author}{\bibfnamefont{E.~J.} \bibnamefont{Torres-Herrera}}
  \bibnamefont{and} \bibinfo{author}{\bibfnamefont{L.~F.}
  \bibnamefont{Santos}}, \bibinfo{journal}{Phys. Rev. E}
  \textbf{\bibinfo{volume}{89}}, \bibinfo{pages}{062110}
  (\bibinfo{year}{2014}{\natexlab{b}}).

\bibitem[{\citenamefont{Torres-Herrera and
  Santos}(2014{\natexlab{c}})}]{Torres2014PRAb}
\bibinfo{author}{\bibfnamefont{E.~J.} \bibnamefont{Torres-Herrera}}
  \bibnamefont{and} \bibinfo{author}{\bibfnamefont{L.~F.}
  \bibnamefont{Santos}}, \bibinfo{journal}{Phys. Rev. A}
  \textbf{\bibinfo{volume}{90}}, \bibinfo{pages}{033623}
  (\bibinfo{year}{2014}{\natexlab{c}}).

\bibitem[{\citenamefont{Torres-Herrera and Santos}(2015)}]{Torres2015}
\bibinfo{author}{\bibfnamefont{E.~J.} \bibnamefont{Torres-Herrera}}
  \bibnamefont{and} \bibinfo{author}{\bibfnamefont{L.~F.}
  \bibnamefont{Santos}}, \bibinfo{journal}{Phys. Rev. B}
  \textbf{\bibinfo{volume}{92}}, \bibinfo{pages}{014208}
  (\bibinfo{year}{2015}).

\bibitem[{\citenamefont{Torres-Herrera
  et~al.}(2015{\natexlab{a}})\citenamefont{Torres-Herrera, T\'avora, and
  Santos}}]{Torres2015BJP}
\bibinfo{author}{\bibfnamefont{E.~J.} \bibnamefont{Torres-Herrera}},
  \bibinfo{author}{\bibfnamefont{M.}~\bibnamefont{T\'avora}}, \bibnamefont{and}
  \bibinfo{author}{\bibfnamefont{L.~F.} \bibnamefont{Santos}},
  \bibinfo{journal}{Braz. J. Phys.} \textbf{\bibinfo{volume}{46}},
  \bibinfo{pages}{239} (\bibinfo{year}{2016}{\natexlab{a}}).

\bibitem[{\citenamefont{Santos et~al.}(2012{\natexlab{a}})\citenamefont{Santos,
  Borgonovi, and Izrailev}}]{Santos2012PRE}
\bibinfo{author}{\bibfnamefont{L.~F.} \bibnamefont{Santos}},
  \bibinfo{author}{\bibfnamefont{F.}~\bibnamefont{Borgonovi}},
  \bibnamefont{and} \bibinfo{author}{\bibfnamefont{F.~M.}
  \bibnamefont{Izrailev}}, \bibinfo{journal}{Phys. Rev. E}
  \textbf{\bibinfo{volume}{85}}, \bibinfo{pages}{036209}
  (\bibinfo{year}{2012}{\natexlab{a}}).

\bibitem[{\citenamefont{Santos et~al.}(2012{\natexlab{b}})\citenamefont{Santos,
  Borgonovi, and Izrailev}}]{Santos2012PRL}
\bibinfo{author}{\bibfnamefont{L.~F.} \bibnamefont{Santos}},
  \bibinfo{author}{\bibfnamefont{F.}~\bibnamefont{Borgonovi}},
  \bibnamefont{and} \bibinfo{author}{\bibfnamefont{F.~M.}
  \bibnamefont{Izrailev}}, \bibinfo{journal}{Phys. Rev. Lett.}
  \textbf{\bibinfo{volume}{108}}, \bibinfo{pages}{094102}
  (\bibinfo{year}{2012}{\natexlab{b}}).

\bibitem[{\citenamefont{Casati et~al.}(1996)\citenamefont{Casati, Chirikov,
  Guarneri, and Izrailev}}]{Casati1996}
\bibinfo{author}{\bibfnamefont{G.}~\bibnamefont{Casati}},
  \bibinfo{author}{\bibfnamefont{B.~V.} \bibnamefont{Chirikov}},
  \bibinfo{author}{\bibfnamefont{I.}~\bibnamefont{Guarneri}}, \bibnamefont{and}
  \bibinfo{author}{\bibfnamefont{F.~M.} \bibnamefont{Izrailev}},
  \bibinfo{journal}{Phys. Lett. A} \textbf{\bibinfo{volume}{223}},
  \bibinfo{pages}{430} (\bibinfo{year}{1996}).

\bibitem[{\citenamefont{Borgonovi et~al.}(2016)\citenamefont{Borgonovi,
  Izrailev, Santos, and Zelevinsky}}]{Borgonovi2016}
\bibinfo{author}{\bibfnamefont{F.}~\bibnamefont{Borgonovi}},
  \bibinfo{author}{\bibfnamefont{F.}~\bibnamefont{Izrailev}},
  \bibinfo{author}{\bibfnamefont{L.~F.} \bibnamefont{Santos}},
  \bibnamefont{and}
  \bibinfo{author}{\bibfnamefont{V.}~\bibnamefont{Zelevinsky}},
  \bibinfo{journal}{Phys. Rep.} \textbf{\bibinfo{volume}{626}},
  \bibinfo{pages}{1} (\bibinfo{year}{2016}).

\bibitem[{\citenamefont{Zelevinsky et~al.}(1996)\citenamefont{Zelevinsky,
  Brown, Frazier, and Horoi}}]{ZelevinskyRep1996}
\bibinfo{author}{\bibfnamefont{V.}~\bibnamefont{Zelevinsky}},
  \bibinfo{author}{\bibfnamefont{B.~A.} \bibnamefont{Brown}},
  \bibinfo{author}{\bibfnamefont{N.}~\bibnamefont{Frazier}}, \bibnamefont{and}
  \bibinfo{author}{\bibfnamefont{M.}~\bibnamefont{Horoi}},
  \bibinfo{journal}{Phys. Rep.} \textbf{\bibinfo{volume}{276}},
  \bibinfo{pages}{85} (\bibinfo{year}{1996}).

\bibitem[{\citenamefont{Santos et~al.}(2011)\citenamefont{Santos, Polkovnikov,
  and Rigol}}]{Santos2011PRL}
\bibinfo{author}{\bibfnamefont{L.~F.} \bibnamefont{Santos}},
  \bibinfo{author}{\bibfnamefont{A.}~\bibnamefont{Polkovnikov}},
  \bibnamefont{and} \bibinfo{author}{\bibfnamefont{M.}~\bibnamefont{Rigol}},
  \bibinfo{journal}{Phys. Rev. Lett.} \textbf{\bibinfo{volume}{107}},
  \bibinfo{pages}{040601} (\bibinfo{year}{2011}).

\bibitem[{\citenamefont{Torres-Herrera and Santos}(2013)}]{Torres2013}
\bibinfo{author}{\bibfnamefont{E.~J.} \bibnamefont{Torres-Herrera}}
  \bibnamefont{and} \bibinfo{author}{\bibfnamefont{L.~F.}
  \bibnamefont{Santos}}, \bibinfo{journal}{Phys. Rev. E}
  \textbf{\bibinfo{volume}{88}}, \bibinfo{pages}{042121}
  (\bibinfo{year}{2013}).

\bibitem[{\citenamefont{Torres-Herrera
  et~al.}(2015{\natexlab{b}})\citenamefont{Torres-Herrera, Kollmar, and
  Santos}}]{TorresKollmar2015}
\bibinfo{author}{\bibfnamefont{E.~J.} \bibnamefont{Torres-Herrera}},
  \bibinfo{author}{\bibfnamefont{D.}~\bibnamefont{Kollmar}}, \bibnamefont{and}
  \bibinfo{author}{\bibfnamefont{L.~F.} \bibnamefont{Santos}},
  \bibinfo{journal}{Phys. Scr. T} \textbf{\bibinfo{volume}{165}},
  \bibinfo{pages}{014018} (\bibinfo{year}{2015}{\natexlab{b}}).

\bibitem[{\citenamefont{He and Rigol}(2012)}]{He2012}
\bibinfo{author}{\bibfnamefont{K.}~\bibnamefont{He}} \bibnamefont{and}
  \bibinfo{author}{\bibfnamefont{M.}~\bibnamefont{Rigol}},
  \bibinfo{journal}{Phys. Rev. A} \textbf{\bibinfo{volume}{85}},
  \bibinfo{pages}{063609} (\bibinfo{year}{2012}).

\bibitem[{\citenamefont{Rigol}(2016)}]{Rigol2015}
\bibinfo{author}{\bibfnamefont{M.}~\bibnamefont{Rigol}},
  \bibinfo{journal}{Phys. Rev. Lett.} \textbf{\bibinfo{volume}{116}},
  \bibinfo{pages}{100601} (\bibinfo{year}{2016}).

\bibitem[{\citenamefont{Wilkinson et~al.}(1997)\citenamefont{Wilkinson,
  Bharucha, C.Fischer, Madison, Morrow, Niu, Sundaram, and
  Raizen}}]{Wilkinson1997}
\bibinfo{author}{\bibfnamefont{S.~R.} \bibnamefont{Wilkinson}},
  \bibinfo{author}{\bibfnamefont{C.~F.} \bibnamefont{Bharucha}},
  \bibinfo{author}{\bibfnamefont{M.}~\bibnamefont{C.Fischer}},
  \bibinfo{author}{\bibfnamefont{K.~W.} \bibnamefont{Madison}},
  \bibinfo{author}{\bibfnamefont{P.~R.} \bibnamefont{Morrow}},
  \bibinfo{author}{\bibfnamefont{Q.}~\bibnamefont{Niu}},
  \bibinfo{author}{\bibfnamefont{B.}~\bibnamefont{Sundaram}}, \bibnamefont{and}
  \bibinfo{author}{\bibfnamefont{M.~G.} \bibnamefont{Raizen}},
  \bibinfo{journal}{Nature (London)} \textbf{\bibinfo{volume}{387}},
  \bibinfo{pages}{575} (\bibinfo{year}{1997}).

\bibitem[{\citenamefont{Brody et~al.}(1981)\citenamefont{Brody, Flores, French,
  Mello, Pandey, and Wong}}]{Brody1981}
\bibinfo{author}{\bibfnamefont{T.~A.} \bibnamefont{Brody}},
  \bibinfo{author}{\bibfnamefont{J.}~\bibnamefont{Flores}},
  \bibinfo{author}{\bibfnamefont{J.~B.} \bibnamefont{French}},
  \bibinfo{author}{\bibfnamefont{P.~A.} \bibnamefont{Mello}},
  \bibinfo{author}{\bibfnamefont{A.}~\bibnamefont{Pandey}}, \bibnamefont{and}
  \bibinfo{author}{\bibfnamefont{S.~S.~M.} \bibnamefont{Wong}},
  \bibinfo{journal}{Rev. Mod. Phys} \textbf{\bibinfo{volume}{53}},
  \bibinfo{pages}{385} (\bibinfo{year}{1981}).

\bibitem[{\citenamefont{Kota}(2001)}]{Kota2001}
\bibinfo{author}{\bibfnamefont{V.~K.~B.} \bibnamefont{Kota}},
  \bibinfo{journal}{Phys. Rep.} \textbf{\bibinfo{volume}{347}},
  \bibinfo{pages}{223} (\bibinfo{year}{2001}).

\bibitem[{\citenamefont{Zangara et~al.}(2013)\citenamefont{Zangara, Dente,
  Torres-Herrera, Pastawski, Iucci, and Santos}}]{Zangara2013}
\bibinfo{author}{\bibfnamefont{P.~R.} \bibnamefont{Zangara}},
  \bibinfo{author}{\bibfnamefont{A.~D.} \bibnamefont{Dente}},
  \bibinfo{author}{\bibfnamefont{E.~J.} \bibnamefont{Torres-Herrera}},
  \bibinfo{author}{\bibfnamefont{H.~M.} \bibnamefont{Pastawski}},
  \bibinfo{author}{\bibfnamefont{A.}~\bibnamefont{Iucci}}, \bibnamefont{and}
  \bibinfo{author}{\bibfnamefont{L.~F.} \bibnamefont{Santos}},
  \bibinfo{journal}{Phys. Rev. E} \textbf{\bibinfo{volume}{88}},
  \bibinfo{pages}{032913} (\bibinfo{year}{2013}).

\bibitem[{not({\natexlab{a}})}]{noteFast}
\bibinfo{note}{Faster than Gaussian decays emerge in full random
  matrices~\cite{Torres2014PRA,Torres2014NJP,Torres2014PRAb} and real systems
  with bimodal density of states~\cite{Torres2014PRAb}.}

\bibitem[{\citenamefont{Khalfin}(1958)}]{Khalfin1958}
\bibinfo{author}{\bibfnamefont{L.~A.} \bibnamefont{Khalfin}},
  \bibinfo{journal}{Zh. Eksp.  Teor. Fiz.} \textbf{\bibinfo{volume}{33}},
  \bibinfo{pages}{1371} (\bibinfo{year}{1958})
  \bibinfo{journal}[{Sov. Phys. JETP} \textbf{\bibinfo{volume}{6}},
  \bibinfo{pages}{1053} (\bibinfo{year}{1958})].
  

\bibitem[{\citenamefont{Nussenzweig}(1961)}]{Nussenzweig1961}
\bibinfo{author}{\bibfnamefont{H.~M.} \bibnamefont{Nussenzweig}},
  \bibinfo{journal}{Nuovo Cim. X} \textbf{\bibinfo{volume}{20}},
  \bibinfo{pages}{694} (\bibinfo{year}{1961}).

\bibitem[{\citenamefont{Ersak}(1969)}]{Ersak1969}
\bibinfo{author}{\bibfnamefont{I.}~\bibnamefont{Ersak}}, \bibinfo{journal}{Yad. Fiz.} \textbf{\bibinfo{volume}{9}}, \bibinfo{pages}{458}
  (\bibinfo{year}{1969})
  \bibinfo{journal}[{Sov. J. Nucl. Phys.} \textbf{\bibinfo{volume}{9}},
  \bibinfo{pages}{263} (\bibinfo{year}{1969})].



\bibitem[{\citenamefont{Fleming}(1973)}]{Fleming1973}
\bibinfo{author}{\bibfnamefont{G.~N.} \bibnamefont{Fleming}},
  \bibinfo{journal}{Il Nuovo Cimento} \textbf{\bibinfo{volume}{16}},
  \bibinfo{pages}{232} (\bibinfo{year}{1973}).

\bibitem[{\citenamefont{Knight}(1977)}]{Knight1977}
\bibinfo{author}{\bibfnamefont{P.}~\bibnamefont{Knight}},
  \bibinfo{journal}{Phys. Lett. A} \textbf{\bibinfo{volume}{61}},
  \bibinfo{pages}{25 } (\bibinfo{year}{1977}).

\bibitem[{\citenamefont{Fonda et~al.}(1978)\citenamefont{Fonda, Ghirardi, and
  Rimini}}]{Fonda1978}
\bibinfo{author}{\bibfnamefont{L.}~\bibnamefont{Fonda}},
  \bibinfo{author}{\bibfnamefont{G.~C.} \bibnamefont{Ghirardi}},
  \bibnamefont{and} \bibinfo{author}{\bibfnamefont{A.}~\bibnamefont{Rimini}},
  \bibinfo{journal}{Rep. Prog. Phys.,} \textbf{\bibinfo{volume}{41}},
  \bibinfo{pages}{587} (\bibinfo{year}{1978}).

\bibitem[{\citenamefont{Sluis and Gislason}(1991)}]{Sluis1991}
\bibinfo{author}{\bibfnamefont{K.~M.} \bibnamefont{Sluis}} \bibnamefont{and}
  \bibinfo{author}{\bibfnamefont{E.~A.} \bibnamefont{Gislason}},
  \bibinfo{journal}{Phys. Rev. A} \textbf{\bibinfo{volume}{43}},
  \bibinfo{pages}{4581} (\bibinfo{year}{1991}).

\bibitem[{\citenamefont{Fock and Krylov}(1947)}]{Fock1947}
\bibinfo{author}{\bibfnamefont{V.}~\bibnamefont{Fock}} \bibnamefont{and}
  \bibinfo{author}{\bibfnamefont{N.}~\bibnamefont{Krylov}},
  \bibinfo{journal}{J. Phys. USSR} \textbf{\bibinfo{volume}{17}},
  \bibinfo{pages}{93} (\bibinfo{year}{1947}).

\bibitem[{\citenamefont{Erd\'elyi}(1956)}]{Erdelyi1956}
\bibinfo{author}{\bibfnamefont{A.}~\bibnamefont{Erd\'elyi}},
  \bibinfo{journal}{J. Soc. Indust. Appr. Math.} \textbf{\bibinfo{volume}{4}},
  \bibinfo{pages}{38} (\bibinfo{year}{1956}).

\bibitem[{\citenamefont{Urbanowski}(2009)}]{Urbanowski2009}
\bibinfo{author}{\bibfnamefont{K.}~\bibnamefont{Urbanowski}},
  \bibinfo{journal}{Eur. Phys. J. D} \textbf{\bibinfo{volume}{54}},
  \bibinfo{pages}{25} (\bibinfo{year}{2009}).

\bibitem[{\citenamefont{Santos and Rigol}(2010{\natexlab{a}})}]{Santos2010PRE}
\bibinfo{author}{\bibfnamefont{L.~F.} \bibnamefont{Santos}} \bibnamefont{and}
  \bibinfo{author}{\bibfnamefont{M.}~\bibnamefont{Rigol}},
  \bibinfo{journal}{Phys. Rev. E} \textbf{\bibinfo{volume}{81}},
  \bibinfo{pages}{036206} (\bibinfo{year}{2010}{\natexlab{a}}).

\bibitem[{\citenamefont{Santos and Rigol}(2010{\natexlab{b}})}]{Santos2010PREb}
\bibinfo{author}{\bibfnamefont{L.~F.} \bibnamefont{Santos}} \bibnamefont{and}
  \bibinfo{author}{\bibfnamefont{M.}~\bibnamefont{Rigol}},
  \bibinfo{journal}{Phys. Rev. E} \textbf{\bibinfo{volume}{82}},
  \bibinfo{pages}{031130} (\bibinfo{year}{2010}{\natexlab{b}}).

\bibitem[{\citenamefont{Chalker and Daniell}(1988)}]{Chalker1988}
\bibinfo{author}{\bibfnamefont{J.~T.} \bibnamefont{Chalker}} \bibnamefont{and}
  \bibinfo{author}{\bibfnamefont{G.~J.} \bibnamefont{Daniell}},
  \bibinfo{journal}{Phys. Rev. Lett.} \textbf{\bibinfo{volume}{61}},
  \bibinfo{pages}{593} (\bibinfo{year}{1988}).

\bibitem[{\citenamefont{Chalker}(1990)}]{Chalker1990}
\bibinfo{author}{\bibfnamefont{J.}~\bibnamefont{Chalker}},
  \bibinfo{journal}{Physica A} \textbf{\bibinfo{volume}{167}},
  \bibinfo{pages}{253 } (\bibinfo{year}{1990}).

\bibitem[{\citenamefont{Kravtsov et~al.}(2011)\citenamefont{Kravtsov, Ossipov,
  and Yevtushenko}}]{Kravtsov2011}
\bibinfo{author}{\bibfnamefont{V.~E.} \bibnamefont{Kravtsov}},
  \bibinfo{author}{\bibfnamefont{A.}~\bibnamefont{Ossipov}}, \bibnamefont{and}
  \bibinfo{author}{\bibfnamefont{O.~M.} \bibnamefont{Yevtushenko}},
  \bibinfo{journal}{J. Phys. A} \textbf{\bibinfo{volume}{44}},
  \bibinfo{pages}{305003} (\bibinfo{year}{2011}).

\bibitem[{\citenamefont{Hsu and d'Auriac}(1993)}]{Hsu1993}
\bibinfo{author}{\bibfnamefont{T.~C.} \bibnamefont{Hsu}} \bibnamefont{and}
  \bibinfo{author}{\bibfnamefont{J.~C. Angl\`es} \bibnamefont{d'Auriac}},
  \bibinfo{journal}{Phys. Rev. B} \textbf{\bibinfo{volume}{47}},
  \bibinfo{pages}{14291} (\bibinfo{year}{1993}).

\bibitem[{\citenamefont{Kudo and Deguchi}(2005)}]{Kudo2005}
\bibinfo{author}{\bibfnamefont{K.}~\bibnamefont{Kudo}} \bibnamefont{and}
  \bibinfo{author}{\bibfnamefont{T.}~\bibnamefont{Deguchi}},
  \bibinfo{journal}{J. Phys. Soc. Jpn.} \textbf{\bibinfo{volume}{74}},
  \bibinfo{pages}{1992} (\bibinfo{year}{2005}).

\bibitem[{\citenamefont{Santos}(2009)}]{Santos2009JMP}
\bibinfo{author}{\bibfnamefont{L.~F.} \bibnamefont{Santos}},
  \bibinfo{journal}{J. Math. Phys} \textbf{\bibinfo{volume}{50}},
  \bibinfo{pages}{095211} (\bibinfo{year}{2009}).

\bibitem[{\citenamefont{Gubin and Santos}(2012)}]{Gubin2012}
\bibinfo{author}{\bibfnamefont{A.}~\bibnamefont{Gubin}} \bibnamefont{and}
  \bibinfo{author}{\bibfnamefont{L.~F.} \bibnamefont{Santos}},
  \bibinfo{journal}{Am. J. Phys.} \textbf{\bibinfo{volume}{80}},
  \bibinfo{pages}{246} (\bibinfo{year}{2012}).

\bibitem[{not({\natexlab{b}})}]{noteXX}
\bibinfo{note}{Due to degeneracies, a high peak at $s=0$ may appear for the
  {$XX$} model and the {$XXZ$} model at the root of unit $\Delta=1/2$
  \cite{Zangara2013}.}

\bibitem[{\citenamefont{Guhr et~al.}(1998)\citenamefont{Guhr,
  Mueller-Gr\"oeling, and Weidenm\"uller}}]{Guhr1998}
\bibinfo{author}{\bibfnamefont{T.}~\bibnamefont{Guhr}},
  \bibinfo{author}{\bibfnamefont{A.}~\bibnamefont{Mueller-Gr\"oeling}},
  \bibnamefont{and} \bibinfo{author}{\bibfnamefont{H.~A.}
  \bibnamefont{Weidenm\"uller}}, \bibinfo{journal}{Phys. Rep.}
  \textbf{\bibinfo{volume}{299}}, \bibinfo{pages}{189} (\bibinfo{year}{1998}).

\bibitem[{\citenamefont{Avishai et~al.}(2002)\citenamefont{Avishai, Richert,
  and Berkovits}}]{Avishai2002}
\bibinfo{author}{\bibfnamefont{Y.}~\bibnamefont{Avishai}},
  \bibinfo{author}{\bibfnamefont{J.}~\bibnamefont{Richert}}, \bibnamefont{and}
  \bibinfo{author}{\bibfnamefont{R.}~\bibnamefont{Berkovitz}},
  \bibinfo{journal}{Phys. Rev. B} \textbf{\bibinfo{volume}{66}},
  \bibinfo{pages}{052416 1} (\bibinfo{year}{2002}).

\bibitem[{\citenamefont{Santos}(2004)}]{Santos2004}
\bibinfo{author}{\bibfnamefont{L.~F.} \bibnamefont{Santos}},
  \bibinfo{journal}{J. Phys. A} \textbf{\bibinfo{volume}{37}},
  \bibinfo{pages}{4723} (\bibinfo{year}{2004}).

\bibitem[{\citenamefont{Dukesz et~al.}(2009)\citenamefont{Dukesz, Zilbergerts,
  and Santos}}]{Dukesz2009}
\bibinfo{author}{\bibfnamefont{F.}~\bibnamefont{Dukesz}},
  \bibinfo{author}{\bibfnamefont{M.}~\bibnamefont{Zilbergerts}},
  \bibnamefont{and} \bibinfo{author}{\bibfnamefont{L.~F.}
  \bibnamefont{Santos}}, \bibinfo{journal}{New J. Phys.}
  \textbf{\bibinfo{volume}{11}}, \bibinfo{pages}{043026 (1}
  (\bibinfo{year}{2009}).

\bibitem[{\citenamefont{Trotzky et~al.}(2008)\citenamefont{Trotzky, Cheinet,
  F\"olling, Feld, Schnorrberger, Rey, Polkovnikov, Demler, Lukin, and
  Bloch}}]{Trotzky2008}
\bibinfo{author}{\bibfnamefont{S.}~\bibnamefont{Trotzky}},
  \bibinfo{author}{\bibfnamefont{P.}~\bibnamefont{Cheinet}},
  \bibinfo{author}{\bibfnamefont{S.}~\bibnamefont{F\"olling}},
  \bibinfo{author}{\bibfnamefont{M.}~\bibnamefont{Feld}},
  \bibinfo{author}{\bibfnamefont{U.}~\bibnamefont{Schnorrberger}},
  \bibinfo{author}{\bibfnamefont{A.~M.} \bibnamefont{Rey}},
  \bibinfo{author}{\bibfnamefont{A.}~\bibnamefont{Polkovnikov}},
  \bibinfo{author}{\bibfnamefont{E.~A.} \bibnamefont{Demler}},
  \bibinfo{author}{\bibfnamefont{M.~D.} \bibnamefont{Lukin}}, \bibnamefont{and}
  \bibinfo{author}{\bibfnamefont{I.}~\bibnamefont{Bloch}},
  \bibinfo{journal}{Science} \textbf{\bibinfo{volume}{319}},
  \bibinfo{pages}{295} (\bibinfo{year}{2008}).

\bibitem[{\citenamefont{Heyl et~al.}(2013)\citenamefont{Heyl, Polkovnikov, and
  Kehrein}}]{Heyl2013}
\bibinfo{author}{\bibfnamefont{M.}~\bibnamefont{Heyl}},
  \bibinfo{author}{\bibfnamefont{A.}~\bibnamefont{Polkovnikov}},
  \bibnamefont{and} \bibinfo{author}{\bibfnamefont{S.}~\bibnamefont{Kehrein}},
  \bibinfo{journal}{Phys. Rev. Lett.} \textbf{\bibinfo{volume}{110}},
  \bibinfo{pages}{135704} (\bibinfo{year}{2013}).

\bibitem[{\citenamefont{Andraschko and Sirker}(2014)}]{Andraschko2014}
\bibinfo{author}{\bibfnamefont{F.}~\bibnamefont{Andraschko}} \bibnamefont{and}
  \bibinfo{author}{\bibfnamefont{J.}~\bibnamefont{Sirker}},
  \bibinfo{journal}{Phys. Rev. B} \textbf{\bibinfo{volume}{89}},
  \bibinfo{pages}{125120} (\bibinfo{year}{2014}).

\bibitem[{\citenamefont{Touchette}(2009)}]{Touchette2009}
\bibinfo{author}{\bibfnamefont{H.}~\bibnamefont{Touchette}},
  \bibinfo{journal}{Phys. Rep.} \textbf{\bibinfo{volume}{478}},
  \bibinfo{pages}{1} (\bibinfo{year}{2009}).

\bibitem[{\citenamefont{Wigner}(1955)}]{Wigner1955}
\bibinfo{author}{\bibfnamefont{E.~P.} \bibnamefont{Wigner}},
  \bibinfo{journal}{Ann. Math.} \textbf{\bibinfo{volume}{62}},
  \bibinfo{pages}{548} (\bibinfo{year}{1955}).

\bibitem[{\citenamefont{Fyodorov et~al.}(1996)\citenamefont{Fyodorov,
  Chubykalo, Izrailev, and Casati}}]{Fyodorov1996}
\bibinfo{author}{\bibfnamefont{Y.~V.} \bibnamefont{Fyodorov}},
  \bibinfo{author}{\bibfnamefont{O.~A.} \bibnamefont{Chubykalo}},
  \bibinfo{author}{\bibfnamefont{F.~M.} \bibnamefont{Izrailev}},
  \bibnamefont{and} \bibinfo{author}{\bibfnamefont{G.}~\bibnamefont{Casati}},
  \bibinfo{journal}{Phys. Rev. Lett.} \textbf{\bibinfo{volume}{76}},
  \bibinfo{pages}{1603} (\bibinfo{year}{1996}).

\bibitem[{\citenamefont{Mirlin et~al.}(1996)\citenamefont{Mirlin, Fyodorov,
  Dittes, Quezada, and Seligman}}]{Mirlin1996}
\bibinfo{author}{\bibfnamefont{A.~D.} \bibnamefont{Mirlin}},
  \bibinfo{author}{\bibfnamefont{Y.~V.} \bibnamefont{Fyodorov}},
  \bibinfo{author}{\bibfnamefont{F.-M.} \bibnamefont{Dittes}},
  \bibinfo{author}{\bibfnamefont{J.}~\bibnamefont{Quezada}}, \bibnamefont{and}
  \bibinfo{author}{\bibfnamefont{T.~H.} \bibnamefont{Seligman}},
  \bibinfo{journal}{Phys. Rev. E} \textbf{\bibinfo{volume}{54}},
  \bibinfo{pages}{3221} (\bibinfo{year}{1996}).

\bibitem[{\citenamefont{Mirlin and Evers}(2000)}]{Mirlin2000}
\bibinfo{author}{\bibfnamefont{A.~D.} \bibnamefont{Mirlin}} \bibnamefont{and}
  \bibinfo{author}{\bibfnamefont{F.}~\bibnamefont{Evers}},
  \bibinfo{journal}{Phys. Rev. B} \textbf{\bibinfo{volume}{62}},
  \bibinfo{pages}{7920} (\bibinfo{year}{2000}).

\bibitem[{\citenamefont{Evers and Mirlin}(2008)}]{Evers2008}
\bibinfo{author}{\bibfnamefont{F.}~\bibnamefont{Evers}} \bibnamefont{and}
  \bibinfo{author}{\bibfnamefont{A.~D.} \bibnamefont{Mirlin}},
  \bibinfo{journal}{Rev. Mod. Phys.} \textbf{\bibinfo{volume}{80}},
  \bibinfo{pages}{1355} (\bibinfo{year}{2008}).

\bibitem[{\citenamefont{Varga}(2002)}]{Varga2002}
\bibinfo{author}{\bibfnamefont{I.}~\bibnamefont{Varga}},
  \bibinfo{journal}{Phys. Rev. B} \textbf{\bibinfo{volume}{66}},
  \bibinfo{pages}{094201} (\bibinfo{year}{2002}).

\bibitem[{\citenamefont{Rigol}(2009{\natexlab{a}})}]{rigol09STATa}
\bibinfo{author}{\bibfnamefont{M.}~\bibnamefont{Rigol}},
  \bibinfo{journal}{Phys. Rev. Lett.} \textbf{\bibinfo{volume}{103}},
  \bibinfo{pages}{100403} (\bibinfo{year}{2009}{\natexlab{a}}).

\bibitem[{\citenamefont{Rigol}(2009{\natexlab{b}})}]{rigol09STATb}
\bibinfo{author}{\bibfnamefont{M.}~\bibnamefont{Rigol}},
  \bibinfo{journal}{Phys. Rev. A} \textbf{\bibinfo{volume}{80}},
  \bibinfo{pages}{053607} (\bibinfo{year}{2009}{\natexlab{b}}).

\bibitem[{\citenamefont{Monnai}(2014)}]{Monnai2014}
\bibinfo{author}{\bibfnamefont{T.}~\bibnamefont{Monnai}}, \bibinfo{journal}{J.
  Phys. Soc. Jpn.} \textbf{\bibinfo{volume}{83}}, \bibinfo{pages}{064001}
  (\bibinfo{year}{2014}).

\bibitem[{\citenamefont{Mazza et~al.}(2016)\citenamefont{Mazza, Stéphan,
  Canovi, Alba, Brockmann, and Haque}}]{Mazza2016}
\bibinfo{author}{\bibfnamefont{P.~P.} \bibnamefont{Mazza}},
  \bibinfo{author}{\bibfnamefont{J.-M.} \bibnamefont{Stéphan}},
  \bibinfo{author}{\bibfnamefont{E.}~\bibnamefont{Canovi}},
  \bibinfo{author}{\bibfnamefont{V.}~\bibnamefont{Alba}},
  \bibinfo{author}{\bibfnamefont{M.}~\bibnamefont{Brockmann}},
  \bibnamefont{and} \bibinfo{author}{\bibfnamefont{M.}~\bibnamefont{Haque}},
  \bibinfo{journal}{J. Stat. Mech.} \textbf{\bibinfo{volume}{2016}},
  \bibinfo{pages}{013104} (\bibinfo{year}{2016}).

\bibitem[{\citenamefont{Polkovnikov}(2011)}]{Polkovnikov2011}
\bibinfo{author}{\bibfnamefont{A.}~\bibnamefont{Polkovnikov}},
  \bibinfo{journal}{Ann. Phys. (N.Y.)} \textbf{\bibinfo{volume}{326}},
  \bibinfo{pages}{486} (\bibinfo{year}{2011}).

\end{thebibliography}

\end{document}